\documentclass[conference]{IEEEtran}
\IEEEoverridecommandlockouts
\usepackage{cite}
\usepackage{amsmath,amssymb,amsfonts}
\usepackage{algorithmic}
\usepackage{graphicx}
\usepackage{textcomp}
\usepackage{xcolor}

\usepackage{caption}
\usepackage{amsmath,epsfig,amssymb,subfigure,bm,dsfont}
\usepackage{lipsum} % 调用跨栏长公式宏包
\usepackage{amsthm}
\usepackage{cite}
\def\BibTeX{{\rm B\kern-.05em{\sc i\kern-.025em b}\kern-.08em
    T\kern-.1667em\lower.7ex\hbox{E}\kern-.125emX}}
\begin{document}

\title{Range-Doppler Information and Doppler Scattering Information in  Multipulse Radar
}

\author{\IEEEauthorblockN{Chao Shi, Dazhuan Xu\textsuperscript{*}, Ying Zhou, Weilin Tu}
\IEEEauthorblockA{\textit{Jiangsu Key Laboratory of Internet of Things and Control Technologies College of Electronic and Information Engineering} \\
\textit{Nanjing University of Aeronautics and Astronautics}\\
Nanjing, China \\
xudazhuan@nuaa.edu.cn}
\and

}

\maketitle

\begin{abstract}
In this paper, the general radar measurement problems of determining range, Doppler frequency and scattering properties parameters are investigated from the viewpoint of Shannon’s information theory. We adopt the mutual information to evaluate the accuracy of the classification and estimation. The range-Doppler information is examined under the condition that the target is of radial velocity. Its asymptotic upper bound and the corresponding entropy error (EE) are further formulated theoretically. Additionally, the Doppler scattering information induced by target’s random motion characteristics is discussed. From the derivation, it is concluded that the Doppler scattering information depends on the eigenvalues of the target scattering correlation matrix. Especially in the case where the pulse interval is larger than target’s coherence time, we can find that the formula of the Doppler scattering information is similar to Shannon’s channel capacity equation, indicating the inherent consistency between the communication theory and radar field. Numerical simulations of these information contents are presented to confirm our theoretical observations. The relationship between the information content and signal-to-noise ratio (SNR) reflects the changes in information acquisition efficiency of a radar system, providing guidance for system designers. 
\end{abstract}

\section{Introduction}\label{sec1}

Doppler radar has been widely used for moving-object detection and target classification. Several kinds of radar systems like moving target indication (MTI) radar, moving target detector (MTD) radar, pulsed Doppler (PD) radar are designed to detect and identify the moving target as well as estimate its relative velocity \cite{richards2005fundamentals,barton1997radar,schleher1991mti,1141412}. \cite{1603402} introduced the micro-Doppler phenomenon in radar. A model of Doppler modulations was developed and the formulas of micro-Doppler induced by targets with micro-motion dynamics were derived therein. Micro-Doppler characteristics were exploited to distinguish among humans, animals, and vehicles in \cite{1174739,1603422,4058282,ISI:000248194500009,4653901}. In synthetic aperture radar (SAR) imaging field, Doppler information is used to obtain the cross-range resolution. At the radar receiver, we can observe the difference in Doppler frequency shift of target’s adjacent scatters, and thus the distribution of the target’s reflectivity can be specified through the Doppler spectrum. The investigations on the range-Doppler imaging algorithm for moving objects can be found in \cite{670330,4102275,4407632}.

Due to the widespread use of the Doppler effect, the estimation of the Doppler frequency is quite significant and a considerable literature exists on this subject \cite{923295,79429,4072241}. However, there are few works on investigating the fundamental links between the derivation of Doppler information content and the estimation process. From the information theory point of view, we can use mutual information to represent the decrease in the a priori uncertainty of the object. Since radar is regarded as an information acquisition system, it is natural and reasonable to apply the mutual information to the radar field for evaluating its estimation performance. 

However, to the best of our knowledge, there are few researches focus on the derivation of the information content for the unknown parameters and adopt it to address the performance analysis of estimation process. For the purpose of introducing this significant feature, we apply the framework of spatial information to multipulse radar. On the one hand, we are concerned about the situation that the target is of radial velocity. In this case, the range-Doppler information can be obtained and then utilized to derive EE, providing guidance to radar system designers from the information theory point of view. Additionally, through the quantity of the information, we are capable of observing the interactions between range and Doppler frequency during the measurement process. Especially when the observation interval is limited and SNR is sufficiently high, we can further conclude that the range information is independent of Doppler information. It is shown that the Doppler scattering information depends on the eigenvalues of the target scattering correlation matrix. In the case where the pulse interval is much larger than target’s coherence time, we can find that the resulting formulation of the Doppler scattering information is consistent with Shannon’s channel capacity equation. This result is quite profound since it demonstrates the fundamental links between the communication theory and radar field. 

This paper is organized as follows. Section 2 provides the theoretical derivation of range-Doppler Information. Its asymptotic upper bound and the corresponding EE are further formulated. In Section 3 the Doppler scattering information is studied and some special cases are discussed.  Section 4 presents the numerical results and, finally, in Section 5, the main results of this paper are discussed and concluded.

\section{Range-Doppler Information }\label{sec2}
\subsection{System Model}
Given $\psi \left( t \right)$ denotes the baseband signal of limited bandwidth $B/2$, the complex envelope of the received signal, scattered by the single moving target, is modelled as
\begin{equation}
	\label{eq1}
	z(t){\rm{ = }}\alpha {e^{j\varphi }}\sum\limits_{m = 0}^{M - 1} {\psi \left( {t - m{T_R} - \tau } \right)} \exp \left( {j2\pi {F_D}t} \right) + w(t)
\end{equation}
where

\qquad $\alpha :$ scattering coefficient of the target;

\qquad $\varphi :$ initial phase;

\qquad $M :$ total number of the transmitted radar pulses;

\qquad ${T_R} :$ pulse repetition interval (PRI);

\qquad ${F_D} :$ Doppler frequency shift;

\qquad $\tau :$ time delay of the round trip between the transmitter and the target. 
\\
In addition, we designate $w\left( t \right)$ as the additive noise, which is modelled as a complex Gaussian zero-mean white random process. 

The sampling sequence can be written as
\begin{equation}
	\label{eq2}
	z\left( n \right)\! = \!\alpha {e^{j\varphi }}\!\sum\limits_{m = 0}^{M - 1} {\psi \left( {n\! -\! m{T_R}B \!- \!x} \right)} \exp \left( {j2\pi {f_d}n} \right) \!+ \!w(n).
\end{equation}

According to the definition of the root mean square bandwidth, we can obtain another expression for the signal bandwidth
\begin{equation}
	\label{eq5}
	\begin{aligned}
		\beta _x^2 &= \frac{{{{\left( {2\pi } \right)}^2}\int_{ - \frac{B}{2}}^{\frac{B}{2}} {{f^2}{{\left| {\Psi \left( f \right)} \right|}^2}df}  - {{\left( {2\pi \int_{ - \frac{B}{2}}^{\frac{B}{2}} {f{{\left| {\Psi \left( f \right)} \right|}^2}df} } \right)}^2}}}{{\int_{ - \frac{B}{2}}^{\frac{B}{2}} {{{\left| {\Psi \left( f \right)} \right|}^2}df} }}\\&
		= \frac{{{\pi ^2}}}{3}{B^2}.
	\end{aligned}
\end{equation}

Since the vector representation is particularly convenient to deal with the noise characteristics, it is favourable for us to introduce here the vector formulation of the received signal. Assume that the number of sampling points is N. This leads to the following observation vector
\begin{equation}
	\label{eq6}
	{\bf{Z}} = \alpha {e^{j\varphi }}{\bf{U}}\left( {x,{f_d}} \right) + {\bf{W}}
\end{equation}
where  $${\bf{Z}} = {\left[ {z\left( { - \frac{N}{2}} \right),...,z\left( {\frac{N}{2} - 1} \right)} \right]^T},$$ and $${\bf{W}} = {\left[ {w\left( { - \frac{N}{2}} \right),...,w\left( {\frac{N}{2} - 1} \right)} \right]^T}.$$ ${\left(  \cdot  \right)^T}$ is the transpose operation. ${\bf{U}}\left( {x,{f_d}} \right)$, a continuous vector function of $x,{f_d}$, denotes the delayed samples of the signal given by (5).
\newcounter{mytempeqncnt}
\begin{figure*}[!t]
	\centering
	\setcounter{mytempeqncnt}{\value{equation}}
	\setcounter{equation}{4}
	\begin{equation}	
		\label{eq5}
		{\bf{U}}\left( {x,{f_d}} \right) = \left( {\begin{array}{*{20}{c}}
				{\sum\limits_{m = 0}^{M - 1} {\sin {\rm{c}}\left( { - \frac{N}{2} - m{T_R}B - x} \right)} \exp \left( { - jN\pi {f_d}} \right)}\\
				{...}\\
				{\sum\limits_{m = 0}^{M - 1} {\sin {\rm{c}}\left( {\frac{N}{2} - 1 - m{T_R}B - x} \right)} \exp \left( { - j\left( {N - 2} \right)\pi {f_d}} \right)}
		\end{array}} \right)
	\end{equation}	
	\hrulefill
	\vspace*{4pt}
\end{figure*}
\setcounter{equation}{\value{mytempeqncnt}}
\addtocounter{equation}{1}   %公式序号加一

Through the observation of the received signal, we are capable of acquiring the desired values of the parameters that characterise the object. According to information theory, the information content is equivalent to the difference of the entropies of the a priori and a posteriori probability distribution, given by
\begin{equation}	
	\label{eq8}
	I\left( {{\bf{Z}};X,{F_d}} \right) = h\left( {X,{F_d}} \right) - h\left( {X,{F_d}|{\bf{Z}}} \right)
\end{equation}	
where
\begin{equation}	
	\label{eq9}
	h\left( {X,{F_d}} \right)\! =\!  - {E_{\rm Z}}\left[ {\int_{\left| {{F_d}} \right|} {\int_{\left| X \right|} {p\left( {x,{f_d}} \right){{\log }_2}p\left( {x,{f_d}} \right)dxd{f_d}} } } \right]\!
\end{equation}	
and
\begin{equation}	
	\label{eq10}
	\begin{aligned}
		&h\left( {X,{F_d}|{\bf{Z}}} \right)\! \\&=\!  - {E_{\rm Z}}\!\left[ {\int_{\left| {{F_d}} \right|} {\int_{\left| X \right|} {p\left( {x,{f_d}|{\bf{Z}}} \right){{\log }_2}p\left( {x,{f_d}|{\bf{Z}}} \right)dxd{f_d}} } } \right].\!
	\end{aligned}
\end{equation}	
$p\left( {x,{f_d}} \right)$ and $p\left( {x,{f_d}|{\bf{Z}}} \right)$ are the a priori and a posterior probability density function, respectively and $E\left[  \cdot  \right]$ denotes the expectation of the random variable.

The a priori distribution of $\alpha$ is formulated by
\begin{equation}	
	\label{eq11}
	p\left( \alpha  \right) = \delta \left( {\alpha  - {\alpha _0}} \right).
\end{equation}	
In the case when frequency is extremely high, a small change of the time delay will result in a huge change of the phase. Therefore we consider $\Phi $ as a variable uniformly distributed in the interval $\left[ {0,2\pi } \right]$, that is
\begin{equation}
	\label{eq12}
	p\left( \varphi  \right) = \frac{1}{{2\pi }}.
\end{equation}
Since the uniform distribution represents a state of the least priori knowledge, it is one of the most general assumptions to make for describing the random properties of the unknown parameters. For this reason, we assume that the a priori distributions of $X$ and ${F_D}$ are uniform on the fixed interval $\left[ {{x_0} - D/2,{x_0} + D/2} \right]$ and $\left[ {{f_{{d_0}}} - \Lambda /2,{f_{{d_0}}} + \Lambda /2} \right]$ respectively near the true value ${x_0}$ and ${f_{{d_0}}}$, namely
\begin{equation}
	\label{eq13}
	p\left( x \right) = \frac{1}{D}
\end{equation}
and
\begin{equation}
	\label{eq14}
	p\left( {{f_d}} \right) = \frac{1}{\Lambda }.
\end{equation}
Note that the values of both $D$ and $\Lambda$ are normalised by $\Delta t = 1/B$ in the sampled signal domain.

\subsection{General Expression}
It is shown from (\ref{eq8}), (\ref{eq9}) and (\ref{eq10}) that, in order to derive the general expression of the information content, the main focus is the formulation of the probability distribution. Since ${\bf{W}}$, as considered previously, is viewed as the complex Gaussian noise samples whose elements are independent identically distributed (i.i.d.) complex Gaussian variables with zero mean and ${N_0}$ variance, we can write its probability density function as
\begin{equation}
	\label{eq15}
	p\left( {\bf{w}} \right) = \frac{1}{{{{\left( {\pi {N_0}} \right)}^{NM}}}}\exp \left( { - \frac{1}{{{N_0}}}{{\left| {\bf{w}} \right|}^2}} \right).
\end{equation}
Applying the form of the received signal as in (\ref{eq6}), we can obtain the probability density function of ${\bf{Z}}$ conditioned on $X,{F_d},\Phi $ and constant ${\alpha _0}$
\begin{equation}
	\label{eq16}
	\!p\left( {{\bf{z}}|x,\!{f_d},\!\varphi } \right)\! =\! \frac{1}{{{{\left( {\pi {N_0}} \right)}^N}}}\!\exp \left( { \!- \!\frac{1}{{{N_0}}}{{\left| {{\bf{z}} \!- \!{\alpha _0}{e^{j\varphi }}{\bf{U}}\left( \!{x,{f_d}}\! \right)} \right|}^2}} \right)\!.\!
\end{equation}
We can derive the a posteriori probability density function as in (15).
\begin{figure*}[!t]
	\centering
	\setcounter{mytempeqncnt}{\value{equation}}
	\setcounter{equation}{14}
	\begin{equation}	
		\label{eq15}
		\begin{aligned}
			p\left( {x,{f_d}|{\bf{z}}} \right) &= \frac{{p\left( {{\bf{z}};x,{f_d}} \right)}}{{\int_{\left| {{F_d}} \right|} {\int_{\left| X \right|} {p\left( {{\bf{z}};x,{f_d}} \right)dxdv} } }}\\&
			= \frac{{\int_0^{2\pi } {\exp \left( {\frac{{2{\alpha _0}}}{{{N_0}}}\Re \left( {{e^{j\varphi }}{{\bf{z}}^H}{\bf{U}}\left( {x,{f_d}} \right)} \right)} \right)d\varphi } }}{{\int_{\left| {{F_d}} \right|} {\int_{\left| X \right|} {\int_0^{2\pi } {\exp \left( {\frac{{2{\alpha _0}}}{{{N_0}}}\Re \left( {{e^{j\varphi }}{{\bf{z}}^H}{\bf{U}}\left( {x,{f_d}} \right)} \right)} \right)d\varphi } dxd{f_d}} } }}.
		\end{aligned}
	\end{equation}
	\setcounter{equation}{28}
	
	\hrulefill
	\vspace*{4pt}
\end{figure*}
\setcounter{equation}{\value{mytempeqncnt}}
\addtocounter{equation}{1}   %公式序号加一

Designating the uncorrelated denominator as a constant coefficient $\mu $ yield
\begin{equation}	
	\label{eq23}
	p\left( {x,{f_d}|{\bf{z}}} \right) = \mu {I_0}\left( {\frac{{2{\alpha _0}}}{{{N_0}}}\left| {{{\bf{z}}^H}{\bf{U}}\left( {x,{f_d}} \right)} \right|} \right).
\end{equation}

Thus we have the general expression of range-Doppler information
\begin{equation}	
	\label{eq25}
	\begin{aligned}
		&I\left( {{\bf{Z}};X,{F_d}} \right) = h\left( {X,{F_d}} \right) - h\left( {X,{F_d}|{\bf{Z}}} \right)\\&
		= {\log _2}D + {\log _2}\Lambda \\&
		+ {E_{\rm Z}}\left[ {\int_{\left| {{F_d}} \right|} {\int_{\left| X \right|} {p\left( {x,{f_d}|{\bf{Z}}} \right){{\log }_2}p\left( {x,{f_d}|{\bf{Z}}} \right)dxd{f_d}} } } \right].
	\end{aligned}
\end{equation}
This information content can be calculated as long as the a posterior probability density in (\ref{eq23}) is obtained. Furthermore, we can figure out its closed asymptotic expression under the specific condition of high SNR, discussed in the following section.

\subsection{Asymptotic Upper Bound}

Substituting the received signal as in (\ref{eq6}) with the actual values of ${x_0}$ and ${f_{{d_0}}}$ into (\ref{eq23}), we can obtain the a posteriori probability density function conditioned on noise
\begin{equation}	
	\label{eq26}
	p\left( {x,{f_d}|{\bf{w}}} \right) \!= \!\mu \!{I_0}\!\left(\! {2{\rho ^2}\left| {{{\bf{U}}^H}\left( {{x_0},{f_{{d_0}}}} \right)\!{\bf{U}}\left( {x,{f_d}} \right) \!+\! \frac{1}{{{\alpha _0}}}{F_w}} \right|} \right)
\end{equation}
where ${\rho ^2} = \alpha _0^2/{N_0}$ denotes the SNR and the noise term ${F_w}$ is defined as
\begin{equation}	
	\label{eq27}
	{F_w} = {e^{ - j{\varphi _0}}}{{\bf{W}}^H}{\bf{U}}\left( {x,{f_d}} \right).
\end{equation}

Then we obtain
\begin{equation}	
	\label{eq29}
	\begin{aligned}
		&p\left( {x,{f_d}|{\bf{w}}} \right) \\&= \mu {I_0}\left( {2{\rho ^2}\left| {{{\bf{U}}^H}\left( {{x_0},{f_{{d_0}}}} \right){\bf{U}}\left( {x,{f_d}} \right)} \right|} \right)\\&
		= \mu {I_0}\left( {2{\rho ^2}\sin c\left( {x - {x_0}} \right)\frac{{\sin \left( {\pi M{T_R}B\left( {{f_d} - {f_{{d_0}}}} \right)} \right)}}{{\sin \left( {\pi {T_R}B\left( {{f_d} - {f_{{d_0}}}} \right)} \right)}}} \right).
	\end{aligned}
\end{equation}
The a posteriori probability density function simplifies to
\begin{equation}	
	\label{eq35}
	\begin{aligned}
		\!p\left( {x,{f_d}|{\bf{w}}} \right)\! =\! \mu \!\exp \left(\! { - \!{\rho ^2}\!M\!\left( {\beta _x^2{{\left( \!{x\! -\! {x_0}} \!\right)}^2} \!+\! \beta _d^2{{\left( {{f_d} \!- \!{f_{{d_0}}}} \right)}^2\!}} \right)} \right)\!
	\end{aligned}
\end{equation}
where some higher order terms about $r - {r_0}$ and ${f_d} - {f_{{d_0}}}$ are neglected since the estimates are considered to locate in the vicinity of true values ${x_0}$ and ${f_{{d_0}}}$ when SNR is high. (\ref{eq35}) demonstrates the fact that that the a posteriori distribution of $X$ and ${F_d}$ is approximately Gaussian near ${x_0}$ and ${f_{{d_0}}}$. Its covariance matrix can be formulated as
\begin{equation}	
	\label{eq36}
	{{\bf{C}}_{X{\bf{,}}{F_d}}} = \left[ {\begin{array}{*{20}{c}}
			{\frac{1}{{2{\rho ^2}\beta _x^2M}}}&0\\
			0&{\frac{1}{{2{\rho ^2}\beta _d^2M}}}
	\end{array}} \right]
\end{equation}
Utilizing the derivation of the Gaussian differential entropy, we can obtain the asymptotic upper bound of the range-Doppler information
\begin{equation}	
	\label{eq37}
	\begin{aligned}
		I\left( {{\bf{Z}};X,{F_d}} \right)&	= {\log _2}\left( {\frac{{D\Lambda {\beta _x}{\beta _d}M{\rho ^2}}}{{\pi e}}} \right).
	\end{aligned}
\end{equation}
It follows from (\ref{eq37}) that the range-Doppler information is determined by the observation interval, root mean square bandwidth, total number of the transmitted radar pulses and SNR. It tells us how these parameters affect the obtainable amount of the information in the measurement process. Furthermore according to (\ref{eq35}) and (\ref{eq36}), it can be observed that the random variables $X$ and ${F_d}$ are independent of each other and thus we can rewrite the range-Doppler information as
\begin{equation}	
	\label{eq38}
	\begin{aligned}
		I\left( {{\bf{Z}};X,{F_d}} \right)& = I\left( {{\bf{Z}};X|{F_d}} \right) + I\left( {{\bf{Z}};{F_d}} \right)\\&
		= I\left( {{\bf{Z}};X} \right) + I\left( {{\bf{Z}};{F_d}} \right)
	\end{aligned}
\end{equation}
where 
\begin{equation}	
	\label{eq39}
	I\left( {{\bf{Z}};X} \right) = {\log _2}\left( {\frac{{D{\beta _x}\sqrt M \rho }}{{\sqrt {\pi e} }}} \right)
\end{equation}
denotes the range information and 
\begin{equation}	
	\label{eq40}
	I\left( {{\bf{Z}};{F_d}} \right) = {\log _2}\left( {\frac{{\Lambda {\beta _d}\sqrt M \rho }}{{\sqrt {\pi e} }}} \right)
\end{equation}
denotes the Doppler information.

\subsection{Entropy Error}
According to the definition of EE in \cite{yan}, we know that the conditional entropy implies the uncertainty of the parameters and can be regarded as a performance metric for the estimation. We formulate EE of range and Doppler frequency as 
\begin{equation}	
	\label{eq41}
	\sigma _{EE}^2{\rm{ = }}\frac{{{2^{2h\left( {X,{F_d}|{\bf{Z}}} \right)}}}}{{{{\left( {2\pi e} \right)}^2}}}.
\end{equation}
Through (\ref{eq25}), we can further derive its relationship with range-Doppler information
\begin{equation}	
	\label{eq42}
	\sigma _{EE}^2{\rm{ = }}\frac{{{D^2}{\Lambda ^2}}}{{{{\left( {2\pi e} \right)}^2}{4^{I\left( {{\bf{Z}};X,{F_d}} \right)}}}}.
\end{equation}
Based on the independence between range and Doppler information as in (\ref{eq38}), EE can be split into two parts
\begin{equation}	
	\label{eq45}
	\begin{aligned}
		\sigma _{EE}^2 &= \frac{{{D^2}}}{{{{\left( {2\pi e} \right)}^2}{4^{I\left( {{\bf{Z}};X} \right)}}}} \cdot \frac{{{\Lambda ^2}}}{{{{\left( {2\pi e} \right)}^2}{4^{I\left( {{\bf{Z}};{F_d}} \right)}}}}\\&
		= \sigma _{xEE}^2 \cdot \sigma _{{f_d}EE}^2
	\end{aligned}
\end{equation}
where $\sigma _{xEE}^2$ and $\sigma _{{f_d}EE}^2$ denote the one-dimensional EE for range and Doppler frequency, respectively.

\section{Doppler Scattering Information}\label{sec3}

Considering the angle between the velocity vector and the radar line of sight is $\theta $, the Doppler frequency shift can be expressed as
\[{F_D} = \frac{{2v\cos \theta }}{\lambda }.\]
Sometimes in practical, the magnitude and the direction of target’s velocity is random, inducing the random change of the scattering phase in received signal. We propose one symbol $s\left( t \right) = \alpha {e^{j\varphi }}\exp \left( {j2\pi {F_D}t} \right)$ to represent the scattering properties of the target, which is regarded as a stationary random process with respect to time $t$. Designating its autocorrelation function as ${R_c}\left( {{\tau _c}} \right)$ and the coherence time as ${T_c}$, we have ${R_c}\left( {{\tau _c}} \right){\rm{ = }}0$ when ${\tau _c} > {T_c}$. The Doppler bandwidth ${B_d}$ is further given by
\begin{equation}	
	\label{eq46}
	{B_d} = \frac{1}{{{T_c}}}.
\end{equation}
Assume that the PRI is large enough so that the sampling points of the pulse waveform has no effect on each other. We sample the received signal $z\left( t \right)$ in (\ref{eq1}) on time $t = m{T_R} + n/B$ and the resulting sequence of the m-th radar pulse simplifies to 
\begin{equation}	
	\label{eq47}
	{{\bf{Z}}_m} = {\bf{U}}\left( x \right){{\rm{s}}_m} + {\bf{W}}
\end{equation}
where ${s_m}$ denotes the observed scattering properties corresponding to the m-th radar pulse and is deemed as invariable in the fast time domain. ${\bf{W}}$ is the noise vector and ${\bf{U}}\left( x \right)$ is the delayed samples of the waveform, given by
\begin{equation}	
	\label{eq48}
	{\bf{U}}\left( x \right){\rm{ = }}{\left[ {\sin c\left( { - \frac{N}{2} - x} \right), \cdots ,sinc\left( {\frac{N}{2} - 1 - x} \right)} \right]^T}.
\end{equation}
Furthermore we can obtain the $MN \times 1$ observation vector of M radar pulses
\begin{equation}	
	\label{eq49}
	\begin{aligned}
		{\bf{Z}} &= {\left[ {\begin{array}{*{20}{c}}
					{{s_1}{\bf{U}}\left( x \right)}& \cdots &{{s_M}{\bf{U}}\left( x \right)}
			\end{array}} \right]^T} + {\bf{W}}\\&
		= {\bf{S}} \otimes {\bf{U}}\left( x \right) + {\bf{W}}
	\end{aligned}
\end{equation}

Through the definition of mutual information, the Doppler scattering information for given range $X = x$ is derived as
\begin{equation}	
	\label{eq50}
	I\left( {{\bf{Z}},\bf{S}}\left | X = x \right. \right) = h\left( {{\bf{Z}}\left| {X = x} \right.} \right) - h\left( {{\bf{Z}}\left| {X = x,\bf{S}} \right.} \right)
\end{equation}

Assuming the scattering properties of the target follow the complex Gaussian distribution, its covariance matrix is given by
\begin{equation}	
	\label{eq52}
	\begin{aligned}
		{{\bf{R}}_{\bf{z}}}&{\rm{ = E}}\left[ {{\bf{Z}}{{\bf{Z}}^{\rm{H}}}} \right]\\&
		= {\rm{E}}\left[ {\left( {{\bf{S}} \otimes {\bf{U}}\left( x \right) + {\bf{W}}} \right){{\left( {{\bf{S}} \otimes {\bf{U}}\left( x \right) + {\bf{W}}} \right)}^{\rm{H}}}} \right]\\&
		= {{\bf{R}}_{\bf{S}}} \otimes {\bf{U}}\left( x \right){{\bf{U}}^{\bf{H}}}\left( x \right) + {N_0}{{\bf{I}}_{{\bf{MN}}}}
	\end{aligned}
\end{equation}
where ${{\bf{R}}_{\bf{S}}}$ is defined as the covariance matrix of scattering sequence ${\bf{S}}$ presented by (36).
\begin{figure*}[!t]
	\centering
	\setcounter{mytempeqncnt}{\value{equation}}
	\setcounter{equation}{35}
	\begin{equation}	
		\label{eq53}
		\begin{aligned}
			{{\bf{R}}_{\bf{S}}}&{\rm{ = E}}\left[ {{\bf{S}}{{\bf{S}}^{\rm{H}}}} \right]\\&
			= \left[ {\begin{array}{*{20}{c}}
					{{R_c}\left( 0 \right)}&{{R_c}\left( {{T_R}} \right)}& \cdots &{{R_c}\left( {\left( {M - 1} \right){T_R}} \right)}\\
					{{R_c}\left( {{T_R}} \right)}&{{R_c}\left( 0 \right)}& \cdots & \cdots \\
					\cdots & \cdots & \cdots &{{R_c}\left( {{T_R}} \right)}\\
					{{R_c}\left( {\left( {M - 1} \right){T_R}} \right)}& \cdots &{{R_c}\left( {{T_R}} \right)}&{{R_c}\left( 0 \right)}
			\end{array}} \right].
		\end{aligned}
	\end{equation}
	\setcounter{equation}{16}
	\hrulefill
	\vspace*{4pt}
\end{figure*}
\setcounter{equation}{\value{mytempeqncnt}}
\addtocounter{equation}{1}   %公式序号加一
Thus the conditional entropy of ${\bf{Z}}$ is formulated as
\begin{equation}	
	\label{eq54}
	h\left( {{\bf{Z}}\left| {X = x} \right.} \right) = {\log _2}\left| {{{\bf{R}}_{\bf{z}}}} \right| + MN{\log _2}\left( {2\pi e} \right).
\end{equation}
The determinant of ${{\bf{R}}_{\bf{z}}}$ can be derived as
\begin{equation}	
	\label{eq57}
	\begin{aligned}
		\left| {{{\bf{R}}_{\bf{z}}}} \right| &= \left| {{\sum _{{{\bf{R}}_{\bf{S}}}}} \otimes {\sum _{{\bf{U}}\left( {\bf{x}} \right){{\bf{U}}^{\bf{H}}}\left( {\bf{x}} \right)}} + {N_0}{{\bf{I}}_{{\bf{MN}}}}} \right|\\&
		= \left( {1{\rm{ + }}\frac{{{\lambda _1}{{\bf{U}}^{\bf{H}}}\left( {\bf{x}} \right){\bf{U}}\left( {\bf{x}} \right)}}{{{N_0}}}} \right) \cdots \left( {1{\rm{ + }}\frac{{{\lambda _M}{{\bf{U}}^{\bf{H}}}\left( {\bf{x}} \right){\bf{U}}\left( {\bf{x}} \right)}}{{{N_0}}}} \right)N_0^{MN}\\&
		= \left( {1{\rm{ + }}\frac{{{\lambda _1}}}{{{N_0}}}} \right) \cdots \left( {1{\rm{ + }}\frac{{{\lambda _M}}}{{{N_0}}}} \right)N_0^{MN}
	\end{aligned}
\end{equation}
where ${\lambda _1},{\lambda _2} \cdots {\lambda _M}$ are M eigenvalues of ${{\bf{R}}_{\rm{\bf{S}}}}$. The Doppler information is given by
\begin{equation}	
	\label{eq58}
	\begin{aligned}
		I\left( {{\bf{Z}},\bf{S}}\left | {X = x} \right. \right) &= h\left( {{\bf{Z}}\left| {X = x} \right.} \right) - h\left( {{\bf{Z}}\left| {X = x,\bf{S}} \right.} \right)\\&
		= {\log _2}\left[ {\left( {\frac{{{\lambda _1}}}{{{N_0}}} + 1} \right) \cdots \left( {\frac{{{\lambda _M}}}{{{N_0}}} + 1} \right)} \right]\\&
		= \sum\limits_{i{\rm{ = }}1}^M {{{\log }_2}\left( {1{\rm{ + }}\frac{{{\lambda _i}}}{{{N_0}}}} \right)} .
	\end{aligned}
\end{equation}
It is shown from (\ref{eq58}) that, the Doppler scattering information depends on the eigenvalues of the target scattering correlation matrix. Especially in the case where the pulse interval is larger than target’s coherence time, namely ${T_R} > {T_c}$, the scattering properties sampled in the slow time are uncorrelated. Then we have
\begin{equation}	
	\label{eq59}
	\begin{aligned}
		{{\bf{R}}_{\rm{\bf{S}}}} = {R_c}\left( 0 \right){{\bf{I}}_{\bf{M}}}\\
		= {E_{\rm{s}}}{{\bf{I}}_{\bf{M}}}.
	\end{aligned}
\end{equation}
In this case, the eigenvalues ${\lambda _1} =  \cdots  = {\lambda _M} = {E_{\rm{s}}}$ equal to the average energy of the scattered signal and the Doppler scattering information of M snapshots can be expressed as
\begin{equation}	
	\label{eq60}
	I\left( {{\bf{Z}},\bf{S}\left| {X = x} \right.} \right) = M{\log _2}\left( {1 + \frac{{{E_s}}}{{{N_0}}}} \right).
\end{equation}
When we adopt the coherence time as PRI, that is
\begin{equation}	
	\label{eq61}
	{T_R} = \frac{1}{{{B_d}}},
\end{equation}
the Doppler scattering information obtained per unit time is given by
\begin{equation}	
	\label{eq62}
	{I_s} = {B_d}{\log _2}\left( {1 + \frac{{{E_s}}}{{{N_0}}}} \right),
\end{equation}
which is proportional to the Doppler spectrum bandwidth and the logarithm of the SNR. It is worth noting that the equation (\ref{eq62}) resembles the formula of Shannon’s channel capacity. In terms of communication theory, we can regard the radar targets as communication sources and Doppler spectrum as the channel bandwidth. This result is quite profound since it demonstrates the fundamental links between the communication theory and radar field.

\section{Numerical Results}\label{sec4}

\begin{figure}
	\centering
	\includegraphics[width=\columnwidth]{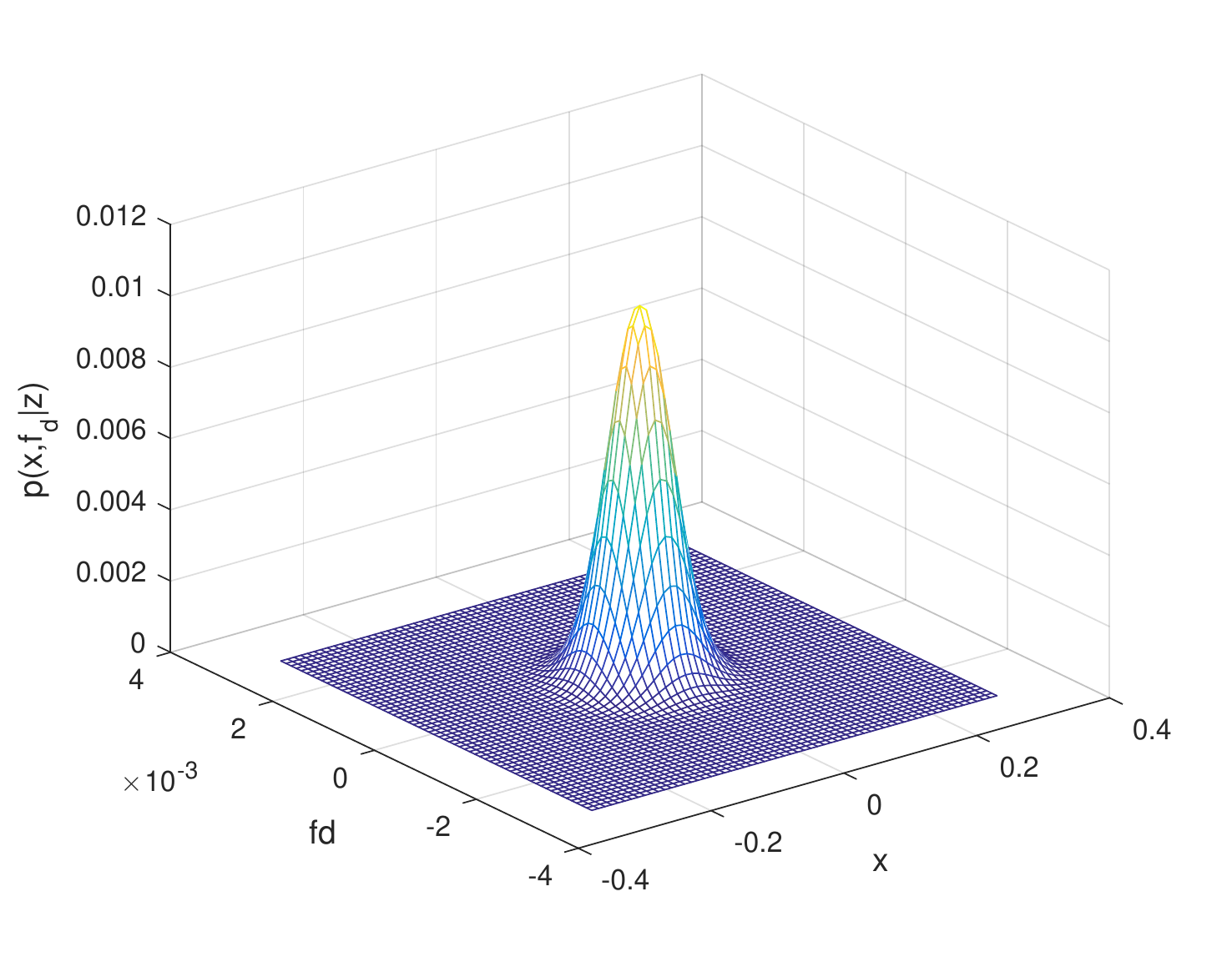}
	\caption{Joint probability density distribution of range and Doppler.}
	\label{fig1}
\end{figure}

In this section, we are going to provide some numerical results to illustrate our theoretical observations of this paper. We consider the number of the sampling points N is 256 and the value of the constant amplitude ${\alpha _0}{\rm{ = }}1$. Fig.\ref{fig1} depicts the joint probability density distribution of range and Doppler frequency at high SNR. It can be seen from this figure that the joint probability density follows two-dimensional Gaussian distribution in the vicinity of the actual values ${x_0} = 0$ and ${f_{{d_0}}} = 0$. 

\begin{figure}
	\centering
	\includegraphics[width=0.9\columnwidth]{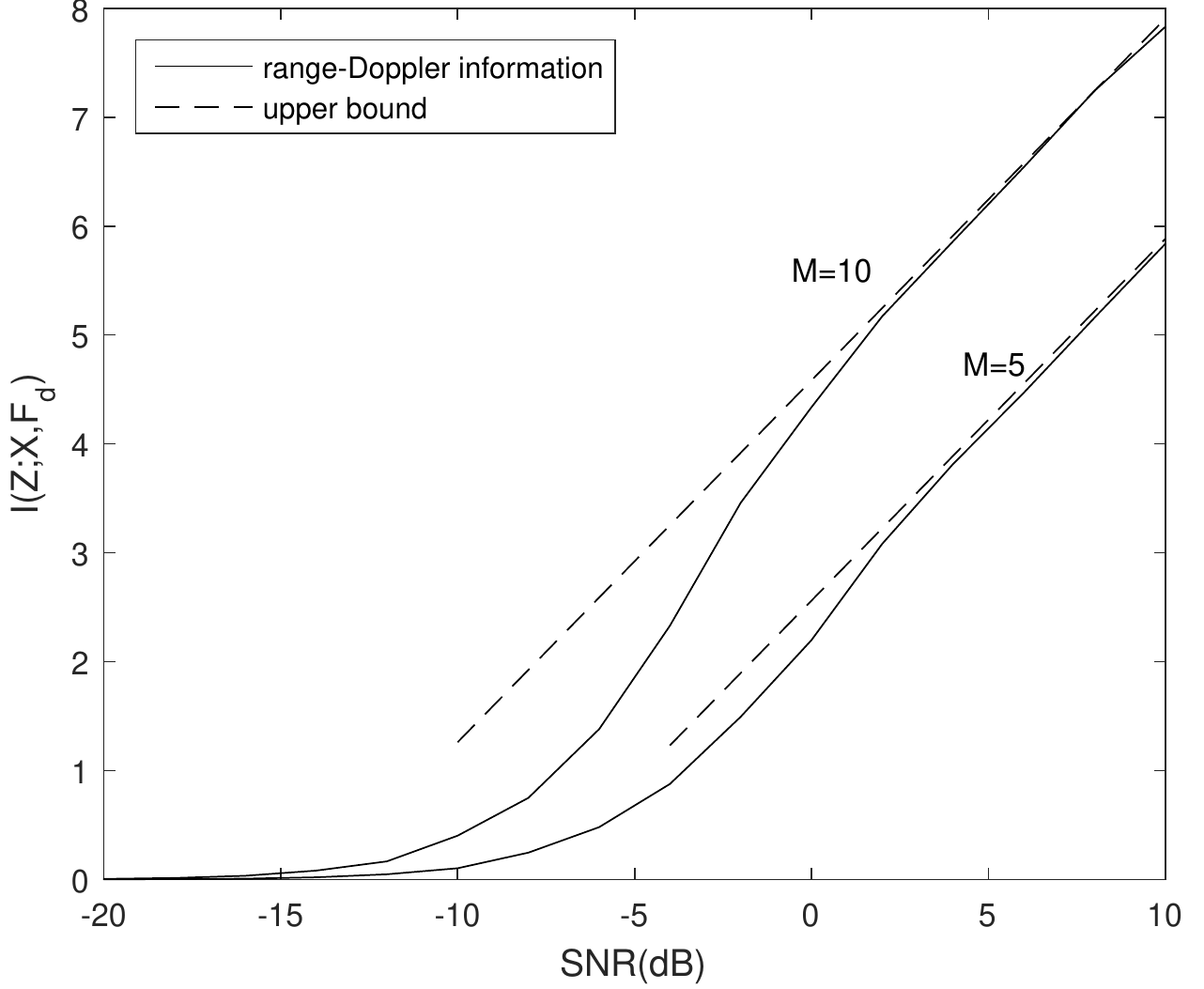}
	\caption{Range-Doppler information and the asymptotic upper bound versus different number of radar pulses.}
	\label{fig2}
\end{figure}

Fig.\ref{fig2} presents the range-Doppler information versus SNR in the case of different number of radar pulses M. Each curve is computed though 100 simulation runs. We can observe that larger M brings greater information content. This is a reasonable phenomenon since when the number of radar pulses increases, we can obtain better Doppler frequency resolution and higher SNR gain form the coherent process, helping us to acquire more information about the unknown parameters.  It also shows that the curve of the information content can be divided into three parts. Firstly we can see that the information content is approximately zero in the case of low SNR. Since the power of Gaussian noise is larger than that of the useful signal, it is difficult to locate the target from the noise. In other words, little information can be obtained in this part. Secondly, the information content grows rapidly with the increase of SNR. We can roughly determine target’s characteristics and the signal energy is being utilized with the maximum efficiency in this part of the SNR region. Finally when SNR is sufficiently high, the amount of the information continues to increase but at a lower speed than before. This phenomenon can be interpreted by the fact that once the approximate characteristics of the target is known without ambiguity, further incoming energy is continually contributing the information which is already partly known. The curves indicates the information acquisition capability of the radar system and thus can provide some guidance for the designers. The derived upper bound of range-Doppler information is also plotted in this figure, verifying the effectiveness of our theoretical analysis in this study.

\begin{figure}
	\centering
	\includegraphics[width=\columnwidth]{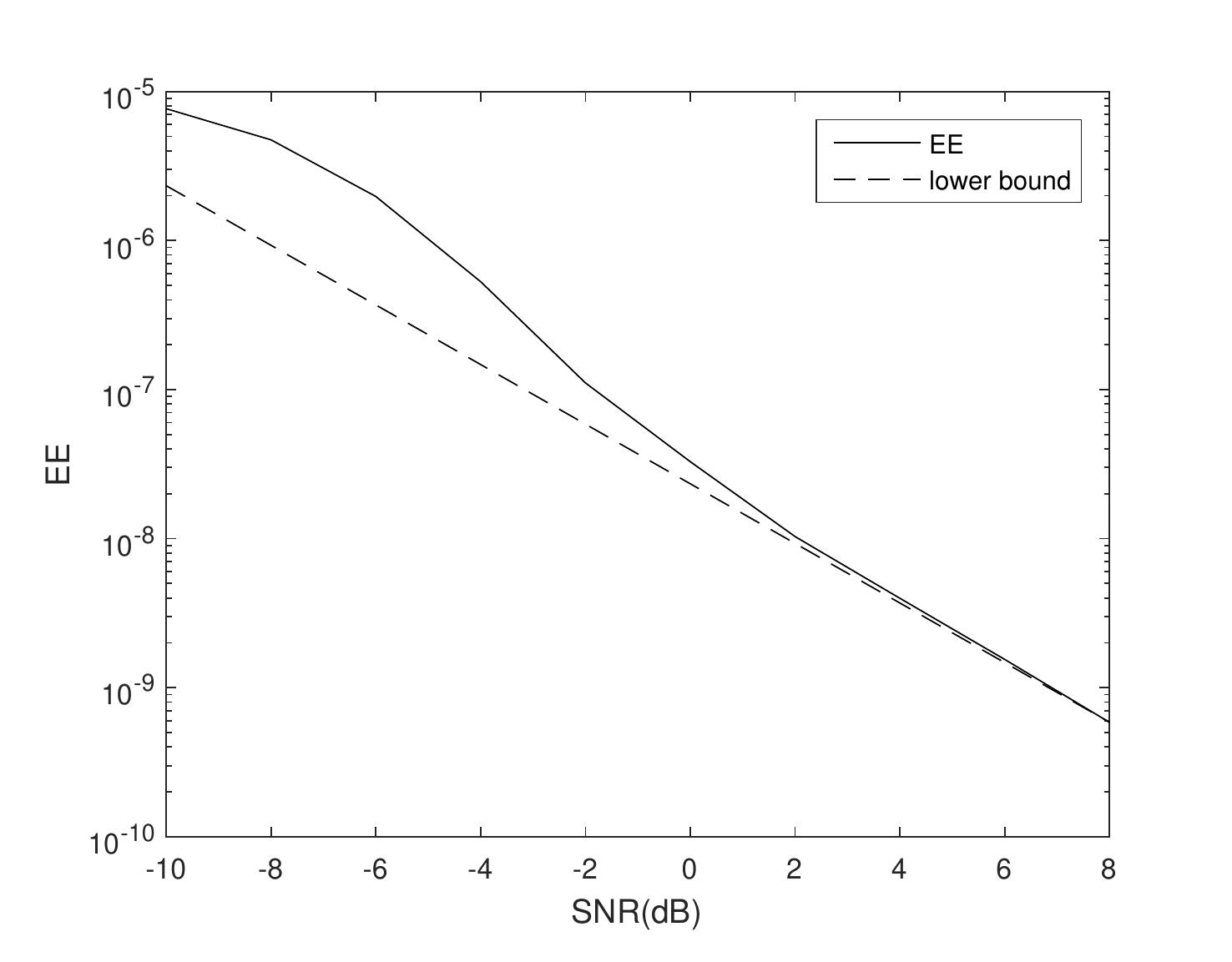}
	\caption{EE and the lower bound versus SNR.}
	\label{fig3}
\end{figure}

Fig.\ref{fig3} depicts EE and its lower bound versus SNR. From (\ref{eq41}), EE can be calculated so long as the conditional entropy is given. It is illustrated from the figure that EE decreases as SNR increases and finally coincides with the lower bound at high SNR. 

\begin{figure}
	\centering
	\includegraphics[width=\columnwidth]{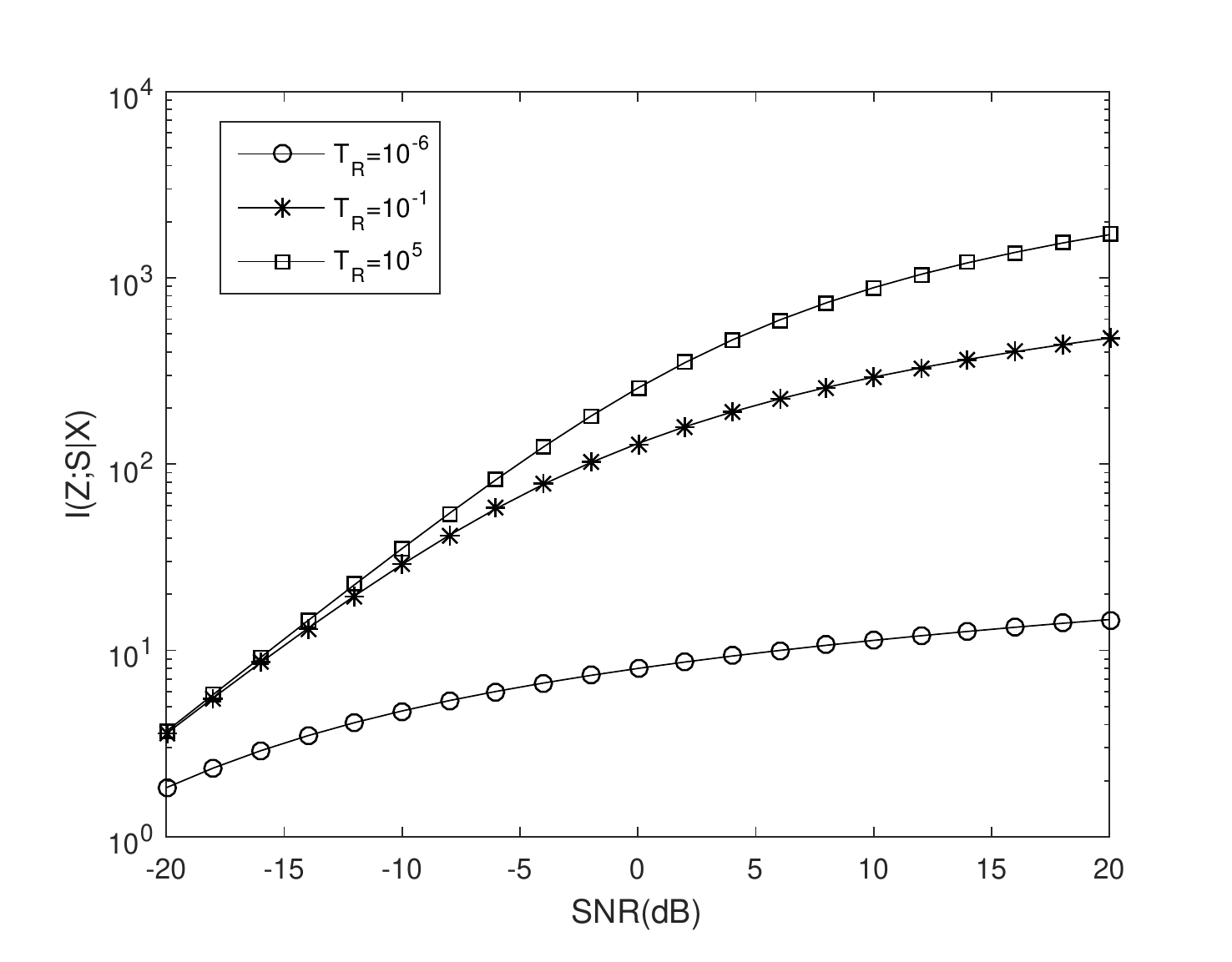}
	\caption{Doppler scattering information versus SNR in the case of different PRI.}
	\label{fig4}
\end{figure}

Fig.\ref{fig4} shows the Doppler scattering information versus SNR in the case of different PRI. In this simulation, a typical Doppler power spectrum in mobile wireless channel, namely Jackes mode, is used. The autocorrelation function of scattering properties is given by
\begin{equation}	
	\label{eq63}
	{R_c}\left( {{\tau _c}} \right) = {E_s}{J_0}\left( {2\pi {f_m}{\tau _c}} \right).
\end{equation}
Here we assume ${f_m} = 1$, ${E_s} = 1$, $M = 256$ and the PRI is set as ${10^{ - 6}}$, ${10^{ - 1}}$, ${10^5}$, respectively. From this figure, we can see that as the PRI increases, the obtained Doppler scattering information increases as well. Specifically, when the value of PRI is quite small, the scattering properties observed from different pulses are considered to be completely correlated. That is to say, the target is stationary during the observation. In this case, ${{\bf{R}}_{\rm{\bf{S}}}}$ is formulated as
\begin{equation}	
	\label{eq64}
	{{\bf{R}}_{\rm{\bf{S}}}}= \left[ {\begin{array}{*{20}{c}}
			{{E_s}}&{{E_s}}& \cdots &{{E_s}}\\
			{{E_s}}&{{E_s}}& \cdots & \cdots \\
			\cdots & \cdots & \cdots &{{E_s}}\\
			{{E_s}}& \cdots &{{E_s}}&{{E_s}}
	\end{array}} \right]
\end{equation}
whose only nonzero eigenvalue is $M{E_s}$. Substituting it into (\ref{eq58}), we have 
\begin{equation}	
	\label{eq65}
	I\left( {{\bf{Z}},\bf{S}}\left | X = x \right. \right) = \log \left( {1 + M{\rho ^2}} \right).
\end{equation}
It demonstrates the fact that the multiple observation for the stationary target can bring us the accumulation of power. Furthermore, as the PRI continues to increase, the target’s scattering properties of different pulses finally become uncorrelated, and the Doppler scattering information is given by (\ref{eq60}). This formula represents the accumulation of the information content. Undoubtedly speaking, such way of accumulation can allow us to learn about the target more efficiently. 

\section{Conclusion}\label{sec5}
In this paper, the general radar measurement problems of determining range, Doppler frequency and scattering properties parameters are investigated from the viewpoint of Shannon’s information theory. We apply the framework of spatial information to multipulse radar. The range-Doppler information is derived and its upper bound is formulated theoretically. It is concluded that the range information is independent of Doppler information when the observation interval is limited and SNR is sufficiently high. Furthermore, based on the conditional entropy, EE is derived and utilized to provide guidance for radar system designers from the information theory point of view. 
Doppler scattering information induced by target’s random motion characteristics is discussed. It is proved that this information content depends on the eigenvalues of the target scattering correlation matrix. Especially in the case where the pulse interval is larger than target’s coherence time, we can find that the formula of the Doppler scattering information is similar to Shannon’s channel capacity equation, indicating the fundamental links between the communication theory and radar field. Numerical simulations are presented to confirm our theoretical observations.

	\bibliographystyle{ieeetr} %% setting the cite style
\bibliography{reference} %% such as \bibiography{iciar}

\end{document}